\documentclass[final]{cvpr}

\usepackage{times}
\usepackage{epsfig}
\usepackage{graphicx}
\usepackage{amsmath}
\usepackage{amssymb}
\usepackage{braket}
\usepackage{mathtools}
\usepackage[title]{appendix}
\DeclarePairedDelimiter\abs{\lvert}{\rvert}%
\DeclarePairedDelimiter\norm{\lVert}{\rVert}%
\usepackage{ragged2e}
\makeatletter
\let\oldabs\abs
\def\abs{\@ifstar{\oldabs}{\oldabs*}}
\let\oldnorm\norm
\def\norm{\@ifstar{\oldnorm}{\oldnorm*}}
\makeatother

\usepackage[pagebackref=true,breaklinks=true,colorlinks,bookmarks=false]{hyperref}

\def\cvprPaperID{8179} 
\def\confYear{CVPR 2021}

\begin{document}

\title{Typical Yet Unlikely and Normally Abnormal: The Intuition Behind High-Dimensional Statistics}

\author{Matthew J. Vowels\\
{\tt\small matthew.vowels@unil.ch}
}

\maketitle
\begin{abstract}
Normality, in the colloquial sense, has historically been considered an aspirational trait, synonymous with ideality. The arithmetic average and, by extension, statistics including linear regression coefficients, have often been used to characterize normality, and are often used as a way to summarize samples and identify outliers. We provide intuition behind the behavior of such statistics in high dimensions, and demonstrate that even for datasets with a relatively low number of dimensions, data start to exhibit a number of peculiarities which become severe as the number of dimensions increases. Whilst our main goal is to familiarize researchers with these peculiarities, we also show that normality can be better characterized with `typicality', an information theoretic concept relating to entropy. An application of typicality to both synthetic and real-world data concerning political values reveals that in multi-dimensional space, to be `normal' is actually to be atypical. We briefly explore the ramifications for outlier detection, demonstrating how typicality, in contrast with the popular Mahalanobis distance, represents a viable method for outlier detection.
\end{abstract}

\section{Introduction}

In a well known United States Air Force (USAF) experiment seeking to identify `the average man', Gilbert Daniels found that out of 4,063 men, not a single one fell within 30\% of the arithmetic sample averages for each of ten physical dimensions simultaneously (which included attributes such as stature, sleeve length, thigh circumference, and so on) \cite{Daniels1952}. Rather than this being an unlikely fluke of the sample, averages, rather than representing the most `normal' attributes in a sample, are actually highly abnormal in the context of multi-dimensional data. Indeed, even though averages may provide seemingly useful baselines for comparison, it is important and perhaps surprising to note that the chance of finding an individual with multiple traits falling close to the average is vanishingly small, particularly as the number of traits increases.

The arithmetic average has been used to represent normality (vis-a-vis abnormality, in the informal/colloquial/non-statistical sense), and is often used both productively and unproductively as a blunt way to characterize samples and outliers. Prior commentary has highlighted the pitfalls associated with the use of the mean as a summary statistic \cite{Speelman2013}; the limitations in relation to its applicability and usefulness of parametric representations (such as the Gaussian) when dealing with real-world phenomena \cite{Micceri1989, Modis2007}; and the societal context \cite{Misztal2002, Comte1976}, surrounding the potentially harmful perception of normality as a ``figure of perfection to which we may progress'' \cite{Hacking1990}. Whilst these commentaries are valuable and important in developing an awareness for what it means to use averages to characterize humankind, they do not provide us with an alternative. They also do not discuss some of the more technical aspects of normality in the context of multiple dimensions and outlier detection, or explain \textit{why} normality, when characterized by the arithmetic average, is so difficult to attain in principle.\footnote{One exception includes work by Kroc and Astivia \cite{Kroc2021} for the determination of scale cutoffs.} Furthermore, the problems associated with the arithmetic average, which is an approximation of an expected value, extend to other expected value based methods, such as regressions, which describe the average value of an outcome for a particular subset of predictors. As such, it is important that researchers familiarize themselves with the limitations of such popular methodologies.

In this paper, our principal aim is to provide intuition for and to familiarize researchers with the peculiar behavior of data in high dimensions. In particular, we discuss averages in the context of multi-dimensional data, and explain why it is that being normal is so abnormal. We touch on some of the peculiarities of multi-dimensional spaces, such as how a high-dimensional sphere has close to zero volume, and how high-dimensional random variables cluster in a thin annulus a finite distance away from the mean. Whilst not the primary focus of this work, we also consider the relevance of these phenomena to outlier detection, and suggest that outliers should not only be considered to include datapoints which lie far from the mean, but also those points close to the mean. The intuition associated with this phenomenon helps to explain why heterogeneity between individuals represents such a great challenge for researchers in the social sciences - humans as high-dimensional phenomena tend only to be normal insofar as they are all abnormal in slightly different ways.

Using information theoretic concepts, we propose an alternative way of characterizing normality and detecting outliers, namely through the concept of `typicality'. We demonstrate the peculiarities as well as the proposed concepts both on idealistic simulated data, as well as data from the `Politics and Views' LISS panel survey \cite{Scherpenzeel2010}.\footnote{A github repository with the Python code used for all analyses, simulations, and plots can be found at \url{https://github.com/matthewvowels1/Typicality}} Finally, we compare the outlier detection performance of typicality with the most common alternative (based on the Mahalanobis distance) and demonstrate it to be a viable alternative. More broadly, we argue that if the average value in a multivariate setting is unlikely, then outlier detection techniques should be able to identify it as such. This means updating our working conceptualization of outliers to include not only points which lie far from the mean (as most outlier detection methods do) but also those points which lie too close to the mean, particularly as the dimensionality of the dataset increases.


\section{Background}

The notion of the mean of a Gaussian, or indeed its finite-sample estimate in the form of the arithmetic average, as representing a `normal person' still holds strong relevance in society and research today. Quetelet did much to popularise the idea \cite{Quetelet1835, Caponi2013}, having devised the much used but also much criticized Body Mass Index for characterizing a person's weight in relation to their height. The average is used as a way to parameterize, aggregate, and compare distributions, as well as to establish bounds for purposes of defining outliers and pathology vis-\`{a}-vis `normality' in individuals. Quetelet's perspective was also shared by Comte, who considered normality to be synonymous with harmony and perfection \cite{Misztal2002}. Even though it is important to recognize the societal and ethical implications of such views, this paper is concerned with the characteristics of multivariate distributions; in particular, those characteristics which help us understand why averages might provide a poor representation of `normality', and what we might consider as an alternative. 

In the past, researchers and commentators (including well-known figures such as Foucault) have levied a number of critiques at the use of averages in psychology and social science  \cite{Speelman2013, Myers2013, Foucault1984, Wetherall1996}. Part of the problem is the over-imposition of the Gaussian distribution on empirical data. The Gaussian has only two parameters, and even if the full probability density function is given, only two pieces of information are required to specify it - the mean (which we treat as equivalent to the arithmetic average) and the variance. Even in univariate cases, the mean can be reductionist, draining the data of nuance and complexity. Many of the developments in statistical methodology have sought to increase the expressivity of statistical models and analyses in order to account for the inherent complexity and heterogeneity associated with psychological phenomena. For example, the family of longitudinal daily diary methods \cite{Bolger2013}, as well as hierarchical models \cite{Raudenbush2002} can be used to capture different levels of variability associated with the data generating process. Alternatively, other methods have sought to leverage techniques from the engineering sciences, such as spectral analysis, in order to model dynamic fluctuations and shared synchrony between partners over time \cite{vowelsplos, Gottman1979}. Machine learning methods provide powerful, data-adaptive function approximation methods for `letting the data speak' \cite{vanderLaan2011} as well as for testing the predictive validity of psychological theories \cite{Vowels2021, Yarkoni2017}, and in the world of big data, comprehensive meta-analyses allow us to paint complete pictures of the gardens of forking paths \cite{Gelman2013, Orben2019}.

Multi-dimensional data exhibit a number of peculiar attributes which concern the use of averages. Assuming one conceives of a `normal person' as having qualities similar to those of a `typical person', we find that the arithmetic average diverges from this conception rather quickly, as the number of dimensions increases. The peculiar attributes start to become apparent in surprisingly low-dimensional contexts (as few as four variables), and become increasingly extreme as dimensionality increases. Understanding these attributes is particularly important because the dimensionality of datasets and analyses is increasing along with the popularity of machine learning. For instance, a machine learning approach identifying important predictors of relationship satisfaction incorporated upwards of 189 variables \cite{Joel2020}, and similar research looking at sexual desire  used around 100 \cite{VowelsSexDes, Joel2017}. Assuming that high-dimensional datasets will continue to be of interest to psychologists, researchers ought to be aware of some of the less intuitive but notable characteristics of such data.

As we will discuss, one domain for which the mean can be especially problematic in multiple-dimensional datasets is outlier detection. In general, outlier detection methods concern themselves with the distance that points lie from the mean. Even methods designed to explore distances from the median are motivated by considerations/difficulties with estimation, and are otherwise based on the assumption that the expected value (or the estimate thereof) provides an object against which to compare datapoints \cite{LeysOutliers}. Unfortunately, and as Daniels' discovered for the USAF, values close to the mean become increasingly unlikely as the number of dimensions increases, making the mean an inappropriate reference for classifying outliers. Indeed, in Daniels' experiment it would have been a mistake to accept anyone close to the average as anything other than an outlier. As we describe later, one \textit{can} successfully summarise a set of datapoints in multiple dimensions in terms of their \textit{typicality}. We later evaluate the performance of a well-known multivariate outlier method (based on the Mahalanobis distance) in terms of its capacity to identify values far from the empirical average as outliers, and compare it against our proposed measure of typicality.


\section{Divergence from the Mean}
This section is concerned with demonstrating some of the un-intuitive aspects of data in higher dimensions. We begin by showing that, as dimensionality increases, the `distance' that a datapoint is from the mean/average increases at a rate of $\sqrt{D}$ where $D$ is the number of dimensions. We then provide a discussion of the ramifications. Finally. we briefly present an alternative geometric view that leads us to the same conclusions. 

\textbf{Notation:} In terms of notation, we denote a datapoint for an individual $i$ as $\mathbf{x}_i$ where $i = \{1, 2, ..., N\}$. The total number of individual datapoints is $N$, and the bold font indicates that the datapoint is a vector (\textit{i.e.} it is multivariate). A single dimension $d$ from individual $i$'s datapoint is given as $x_{i,d}$, where $d \in (\mathbb{Z}^{+})^D$, where $D$ is the total number of dimensions.

\begin{figure}[!t]
\centering
\includegraphics[scale=0.3]{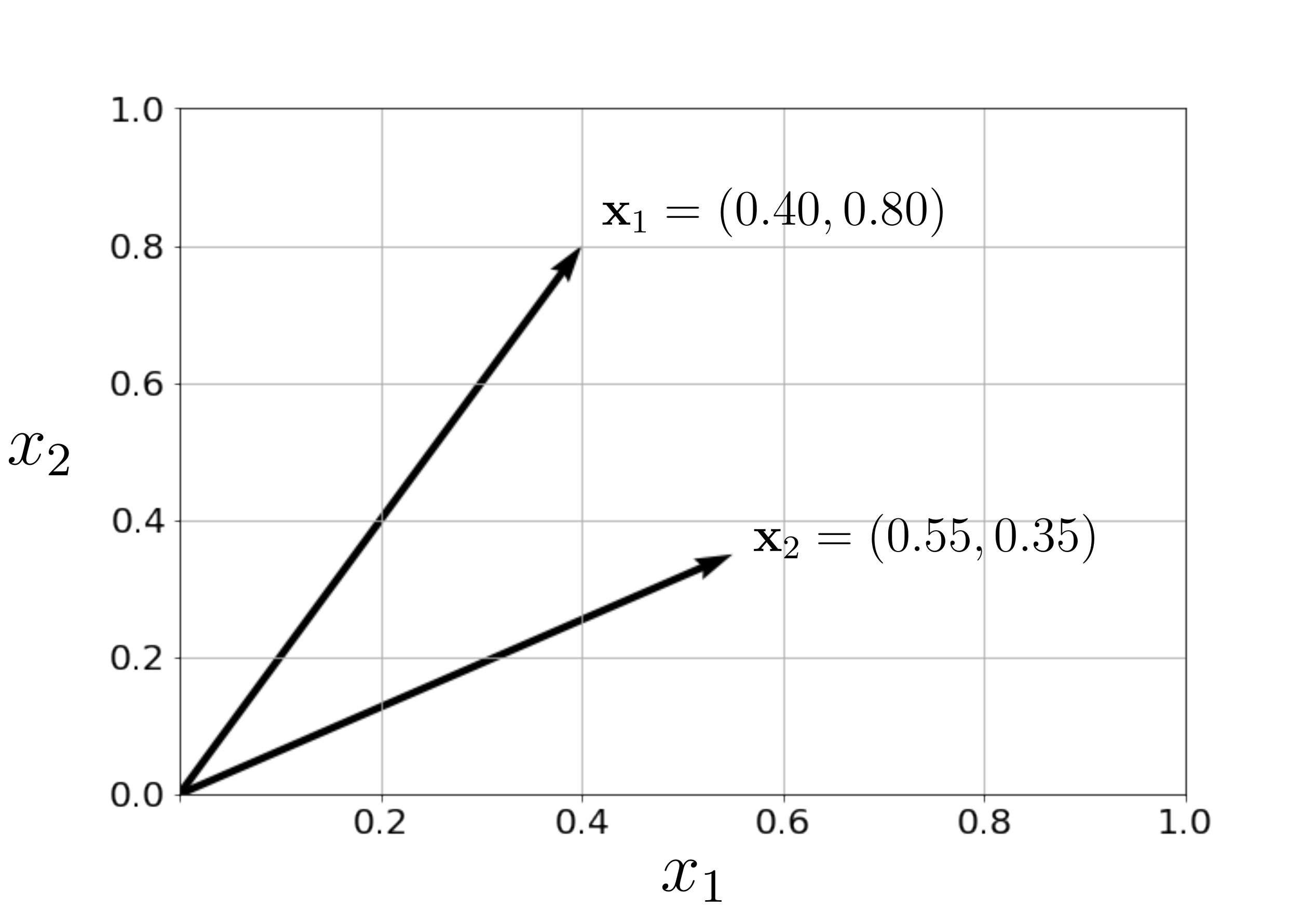}
\newline
\caption{Two samples in two-dimensional space, with their corresponding coordinates.}
\label{fig:vectors}
\end{figure}

\subsection{Gaussian Vectors in High Dimensions}

Let us begin in familiar territory - for a multivariate distribution with independently and identically distributed (\textit{i.i.d}) Gaussian variables, the probability density function for \textit{each dimension} may be expressed as:

\begin{equation}
    p(x_{i,d}) = \frac{1}{\sqrt{2 \pi \sigma^2_d}}e^{-\frac{(x_{i,d}-\mu_{d})^2}{2\sigma_d^2}}
\end{equation}

where $\mu_d$ and $\sigma_d$ represent the mean and standard deviation of dimension $d$, respectively. Each multivariate datapoint $\mathbf{x}_i$ may be considered as a vector in this $D$-dimensional space. An example of two datapoints drawn from a two-dimensional/bivariate version of this distribution (\textit{i.e.}, $D=2$), is shown in Figure \ref{fig:vectors}. In this figure, the values of these two random samples are $\mathbf{x}_1 = (0.4, 0.8)$ and $\mathbf{x}_2 = (0.55, 0.35)$. Assuming that these datapoints are drawn from a distribution with a mean of 0 and a variance of 1 for all dimensions (\textit{i.e.}, $\mathcal{N}(\mu_d = 0, \sigma^2_d = 1) \; \;  \forall \; \; d$), then we can compute the distance these datapoints fall from the mean $\boldsymbol{\mu} = \mathbf{0}$ using the squared Euclidean distance (see Eq. \ref{eq:norm}),

\begin{equation}
    ||\mathbf{x}_i||_2^2 = \sum_{d=1}^D x_{i, d}^2
    \label{eq:norm}
\end{equation}

Here, we use the subscript $d$ to index the dimension of the multidimensional datapoint $\mathbf{x}_i$. Importantly, note that the \textit{squared} Euclidean norm closely resembles the expression for sample variance for a certain dimension $d$ (Eq.~\ref{eq:var}):

\begin{equation}
    \overline{\mbox{Var}}(x_d) = \frac{1}{N}\sum_{i=1}^N x_{i, d}^2
    \label{eq:var}
\end{equation}

For the two example vectors in Figure \ref{fig:vectors}, taking the square root of the values derived using Eq.~\ref{eq:norm}, the distances from the mean are $||\mathbf{x}_1||_2 = 0.8$ and $||\mathbf{x}_2||_2 = 0.3$. 

In other words, the variance of a sample is closely related to the distance that each sample is expected to fall from the mean. Note that, when computing the variance for a particular variable or dimension, we sum across datapoints $i$, rather than dimensions $d$. Secondly, and more importantly, the variance contains a normalization term $1/N$, whereas the expression for the norm does not. By consequence of this absent normalization term, the expected squared distance of each datapoint from the mean will grow with increasing dimensionality. In this example, we know that the variance of our distribution $\sigma^2_d = 1$ for both dimensions, and as such, it is trivial to show that each individual dimension $d$ will have an average/expected length equal to one. Without the normalization term (\textit{i.e.}, $1/D$), this means that the expected squared length of the vectors grows in proportion to the number of dimensions.\footnote{This is simply the result of summing more values together, without accounting for the number of dimensions. This is intentional - to compute the overall distance a sample lies from the mean, it is necessary to sum across all possible dimensions/directions \textit{without} adjusting for the number of dimensions.} Alternatively, taking a square root, we can say that the expected distance that samples fall from the mean increases proportional to the square-root of the dimensionality of the distribution. More concretely: $\mathbb{E} \left[ ||\mathbf{x}||_2 \right]\propto \sqrt{D}$ \cite{Vershynin2019}.

\begin{figure}[!t]
\centering
\includegraphics[scale=0.3]{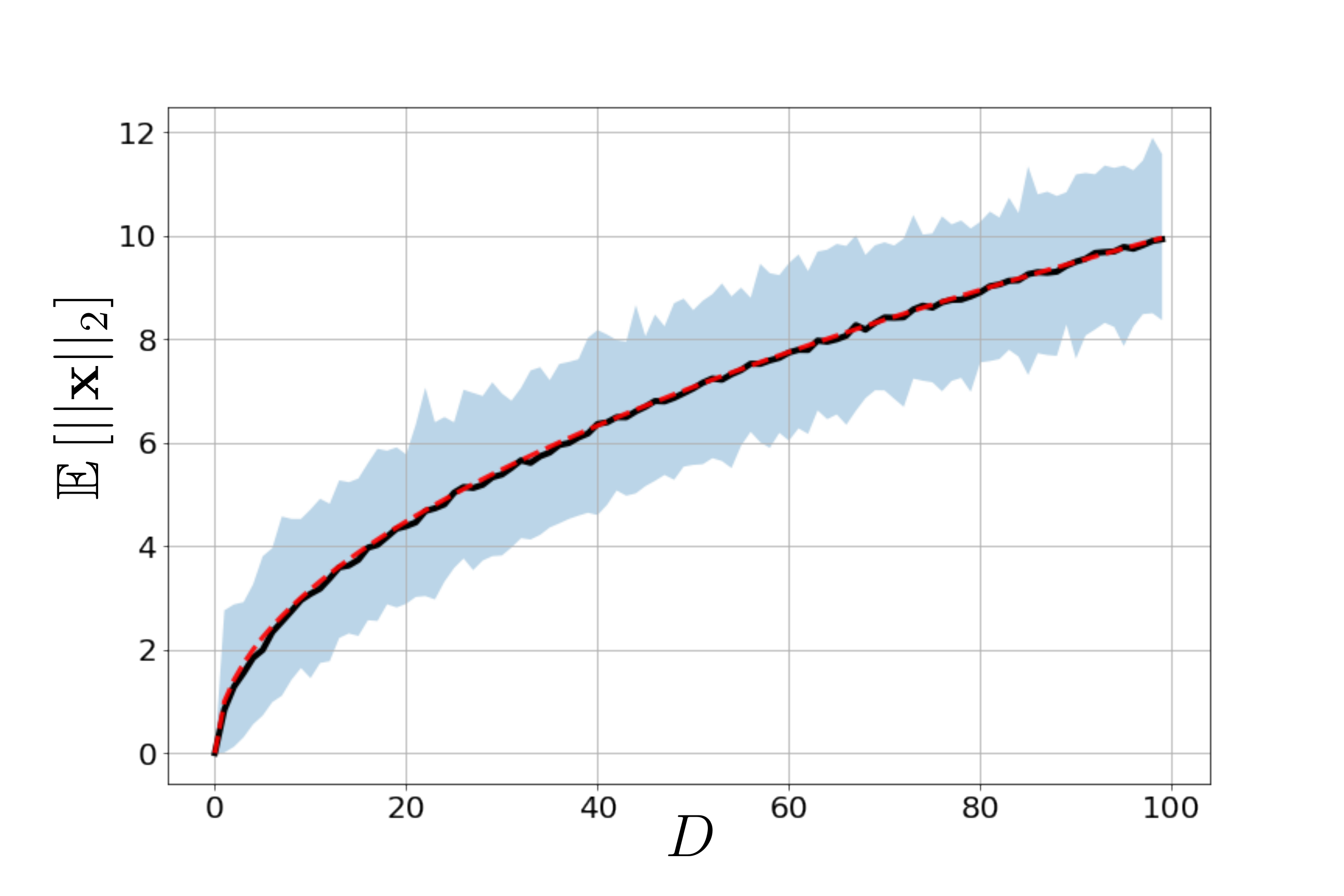}
\newline
\caption{The red dashed curve is simple $\sqrt{D}$, whilst the black curve is a simulated estimate of the expected lengths, calculated over 200 datapoints, for increasing dimensionality $D$. The blue interval represents the 1-99\% percentiles. }
\label{fig:exp_lengths}
\end{figure}

This can of course also be verified in simulation, and Figure \ref{fig:exp_lengths} shows both the analytical as well as sample estimates for the average length of the vectors as the number of dimensions increases. The intervals are defined by the 1st and 99th percentiles. Each approximation to the expectation is taken over a sample size of 200 datapoints. The dashed red curve depicts the $\sqrt{D}$ relationship, and the black simulated curve is a direct (albeit noisier, owing to the fact that this curve is simulated) overlay. This should start to remind us of Daniels' experience when working for the USAF - he found that out of 4,063 people, not a single one of them fell within 30\% of the mean over ten variables. Indeed, if any had done, we should consider labelling them as outliers, in spite of the fact that most existing outlier detection methods are only sensitive to points which lie \textit{far} from the mean.

\begin{figure}[!t]
\centering
\includegraphics[scale=0.3]{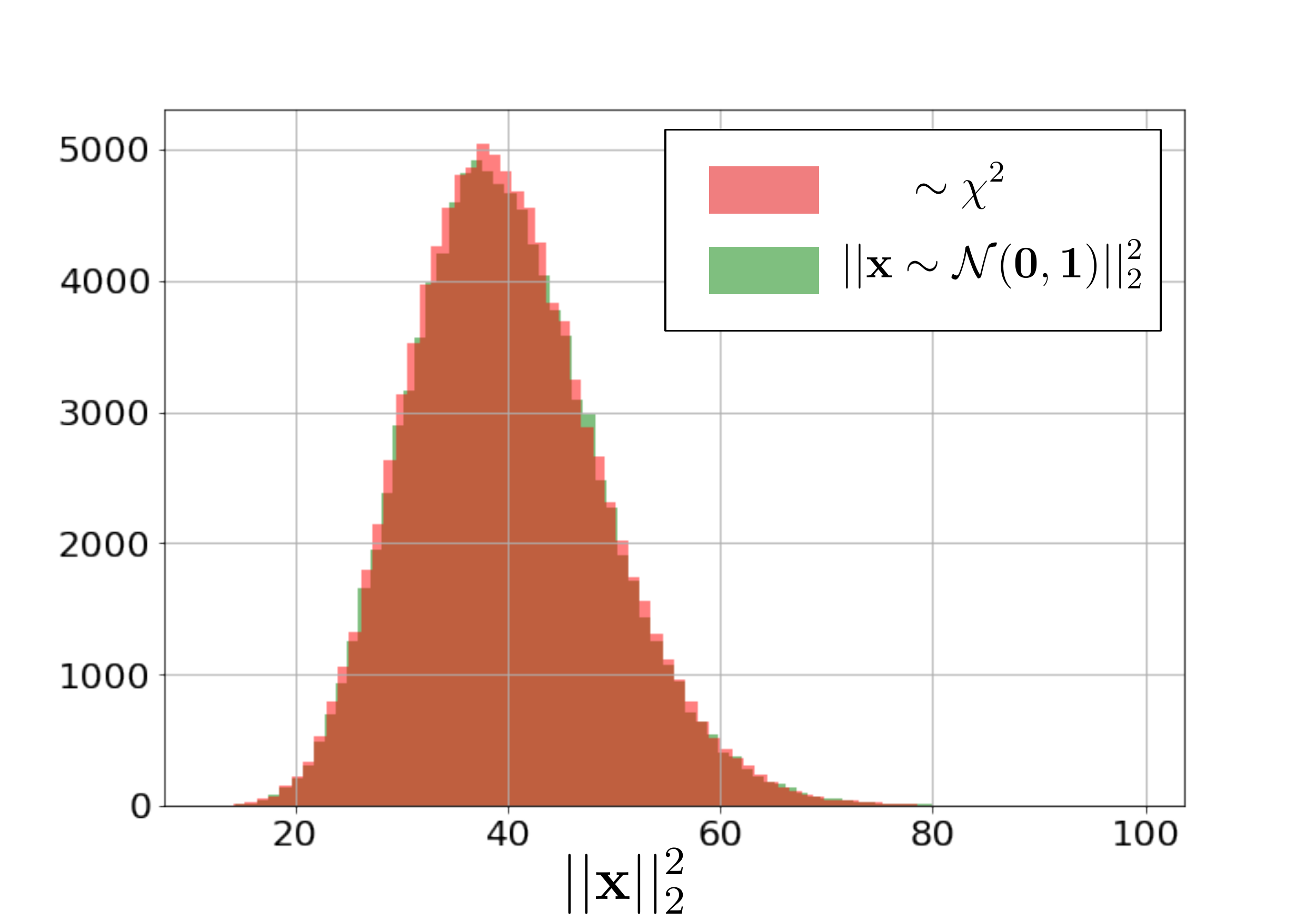}
\newline
\caption{For $D=40$ these histograms show the distributions of 10,000 datapoints sampled from a $\chi^2$ distribution (red) and the sums of squared distances $||\mathbf{x}||^2_2$.}
\label{fig:histograms}
\end{figure}

The implications of this are important to understand. Whilst we know that each variable $x_d$ has an expected value of zero and a variance of one, the expected length of a datapoint across all $D$ dimensions/variables grows in proportion to the square root of the number of variables. Dieleman \cite{Dieleman2020} summarised this informally when they observed that ``if we sample lots of vectors from a 100-dimensional standard Gaussian, and measure their radii, we will find that just over 84\% of them are between 9 and 11, and more than 99\% are between 8 and 12. Only about 0.2\% have a radius smaller than 8!'' In other words, the expected location of a datapoint in $D$-dimensional space moves further and further away from the mean as the dimensionality increases. 

\begin{figure}[!t]
\centering
\includegraphics[scale=0.4]{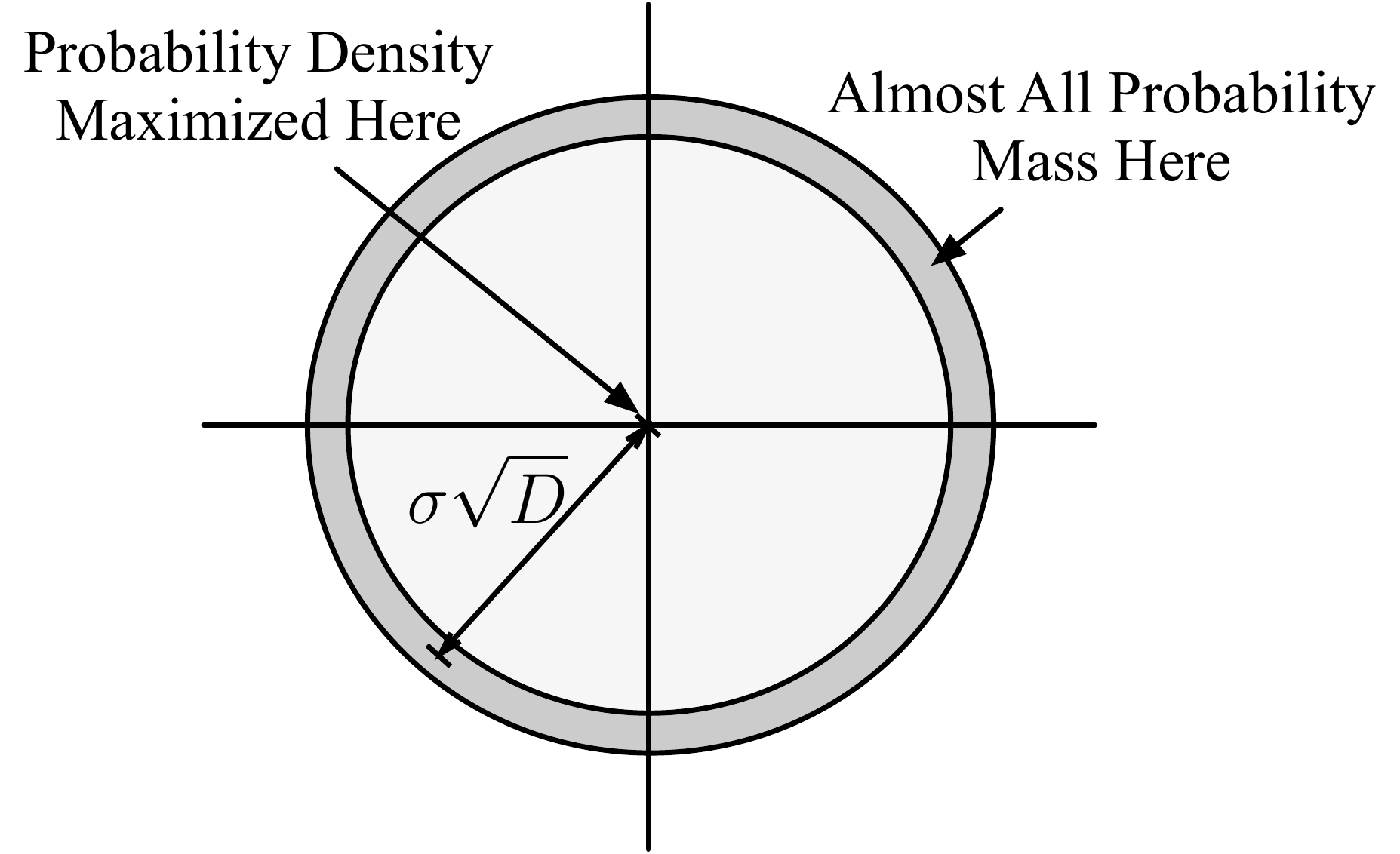}
\newline
\caption{The plot illustrates how, in high-dimensions, the probability mass is located in a thin annulus at a distance $\sigma\sqrt{D}$ from the average (in the text, we assume $\sigma=1$), despite the mean representing the location which maximizes the probability density. Adapted from \cite{mackay}.}
\label{fig:Mackay}
\end{figure}

It can also be shown that such high-dimensional Gaussian random variables are distributed uniformly on the (high-dimensional) sphere with a radius of $\sqrt{D}$, and grouped in a thin annulus \cite{Stein2020, Vershynin2019}.\footnote{See also Gaussian Annulus Theorem \cite{Blum2020}.} The uniformity tells us the direction of these vectors (\textit{i.e.}, their location on the surface of this high-dimensional sphere) is arbitrary, and the squared distances, or radii, are Chi-squared distributed (it is well known that the Chi-squared distribution is the distribution of the sum of squares of $D$ independent and identically distributed Gaussian variables). The distribution of distances (vis-\`{a}-vis the squared distances) is therefore Chi-distributed.  Figure \ref{fig:histograms} compares samples from a Chi-squared distribution against the distribution of 10,000 squared vector lengths. Altogether, this means that in high-dimensions, (a) it is unlikely to find datapoints anywhere close to the average (even though the region close to the mean represents the one with the highest likelihood, the probability is nonetheless negligible), (b) randomly sampled vectors are unlikely to be correlated (of course, in expectation the correlation will be zero because the dimensions of the Gaussian from which they were sampled are independent), and (c) randomly sampled vectors have lengths that are close to the expected length which increases at a rate $\sqrt{D}$. As such, the datapoints tend to cluster in a subspace which lies at a fixed radius from the mean (we will later refer to this subspace as the \textit{typical set}). This is summarized graphically in Figure \ref{fig:Mackay}.

\begin{figure}[!t]
\centering
\includegraphics[scale=0.35]{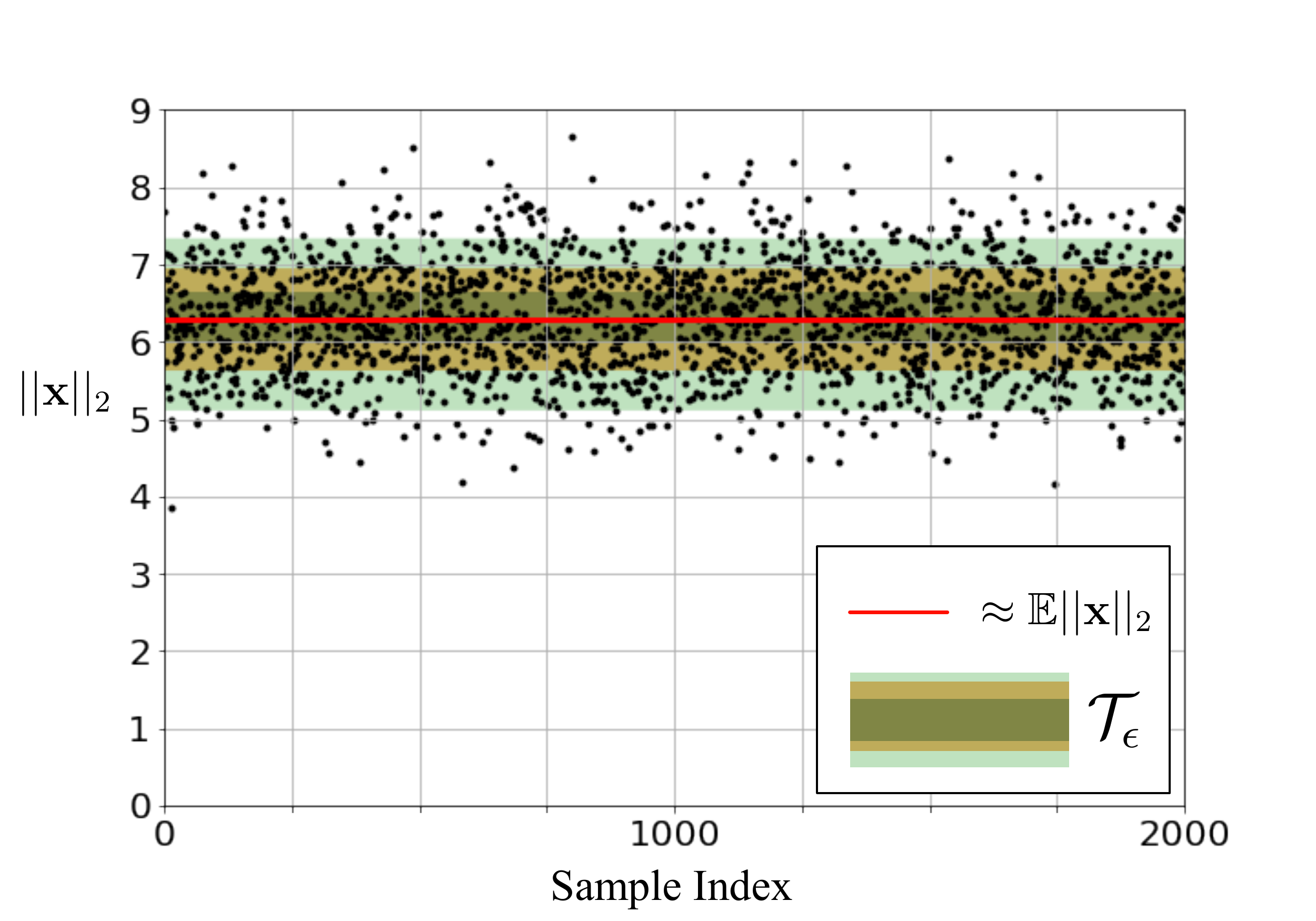}
\newline
\caption{A scatter plot showing the lengths of 2,000 vectors sampled from a 40-dimensional Gaussian. Red line shows the average vector length, and the green intervals depict the size of the typical set for different values of $\epsilon$. Note that the mean (0,0) is nowhere near the distribution of norms or the typical set. }
\label{fig:typicals}
\end{figure}

\begin{figure}[!t]
\centering
\includegraphics[scale=0.3]{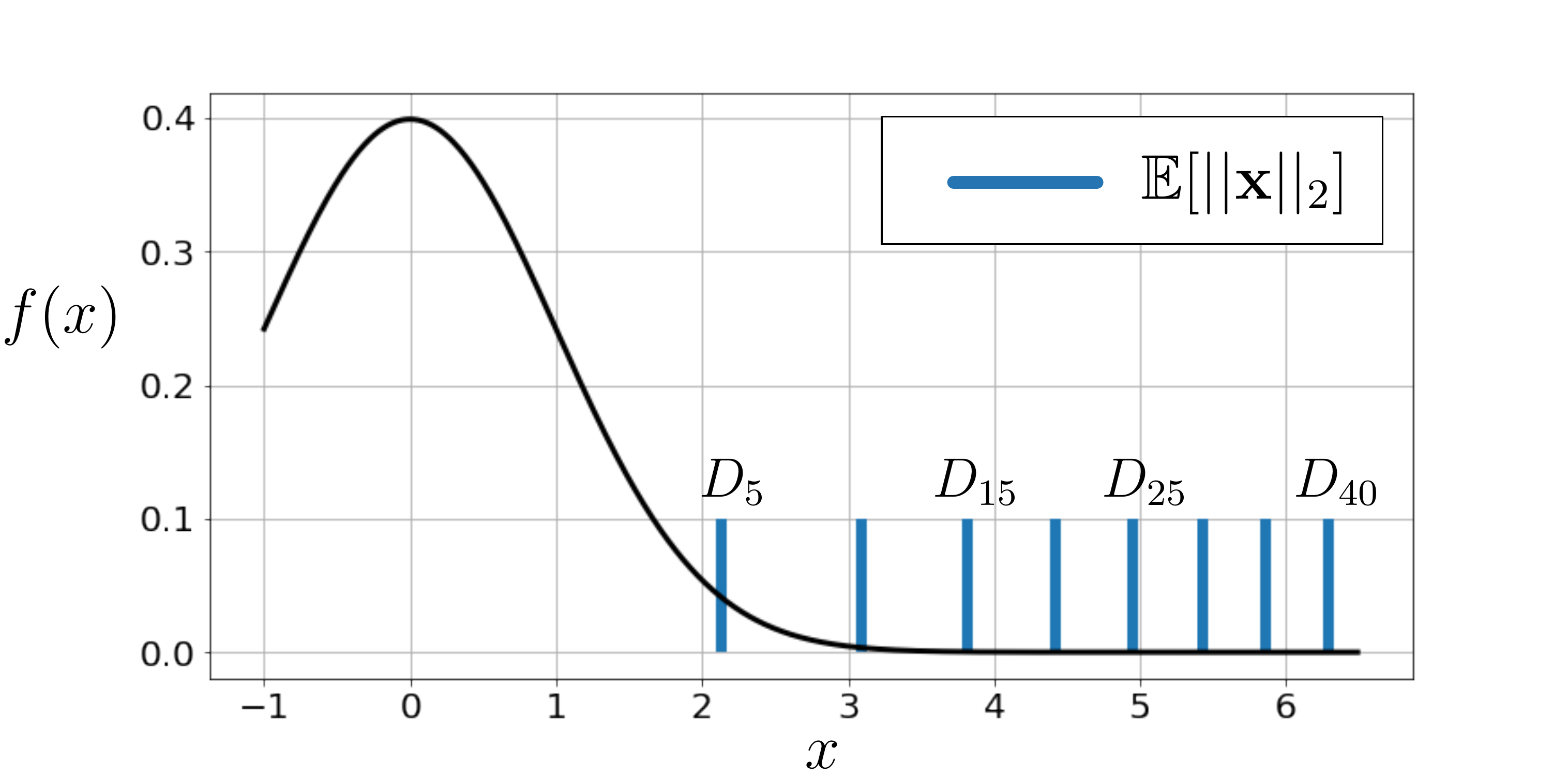}
\newline
\caption{This plot shows the location of the expected lengths of vectors of different dimensionality in relation to the standard normal in one dimension. It can be seen that even at $D=5$, the expected length is over two standard deviations from the mean. }
\label{fig:gauss}
\end{figure}

It is important that researchers understand that while the mean of such a high-dimensional Gaussian represents the value which minimizes the sums of squared distances (and is therefore the estimate which maximises the likelihood), the majority of the probability mass is actually \textit{not} located around this point. As such, even though a set of values close to the mean represents the most likely in terms of its probability of occurrence, the magnitude of this probability is negligible, and most points fall in a space around $\sqrt{D}$ away from the mean. Figure \ref{fig:typicals} depicts the lengths of 2000 vectors sampled from a 40-dimensional Gaussian - they are nowhere close to the origin. Another way to visualize this is to plot the locations of the expected lengths for different dimensionalities on top of the curve for $\mathcal{N}(0,1)$, and this is shown in Figure \ref{fig:gauss}. In terms of the implications for psychological data - datasets which involve high numbers of variables are likely to comprise individuals who are similar only insofar as they appear to be equally `abnormal', at least insofar as a univariate characterization of normality (e.g. the mean across the dimensions) is a poor one when used across multiple dimensions. Indeed, if an individual \textit{does} possess characteristics close to the mean or the mode across multiple dimensions, they could reasonably be considered to be \textit{outliers}. We will consider outlier detection more closely in a later section.

\subsection{An Alternative Perspective}

\begin{figure}[!t]
\centering
\includegraphics[scale=0.3]{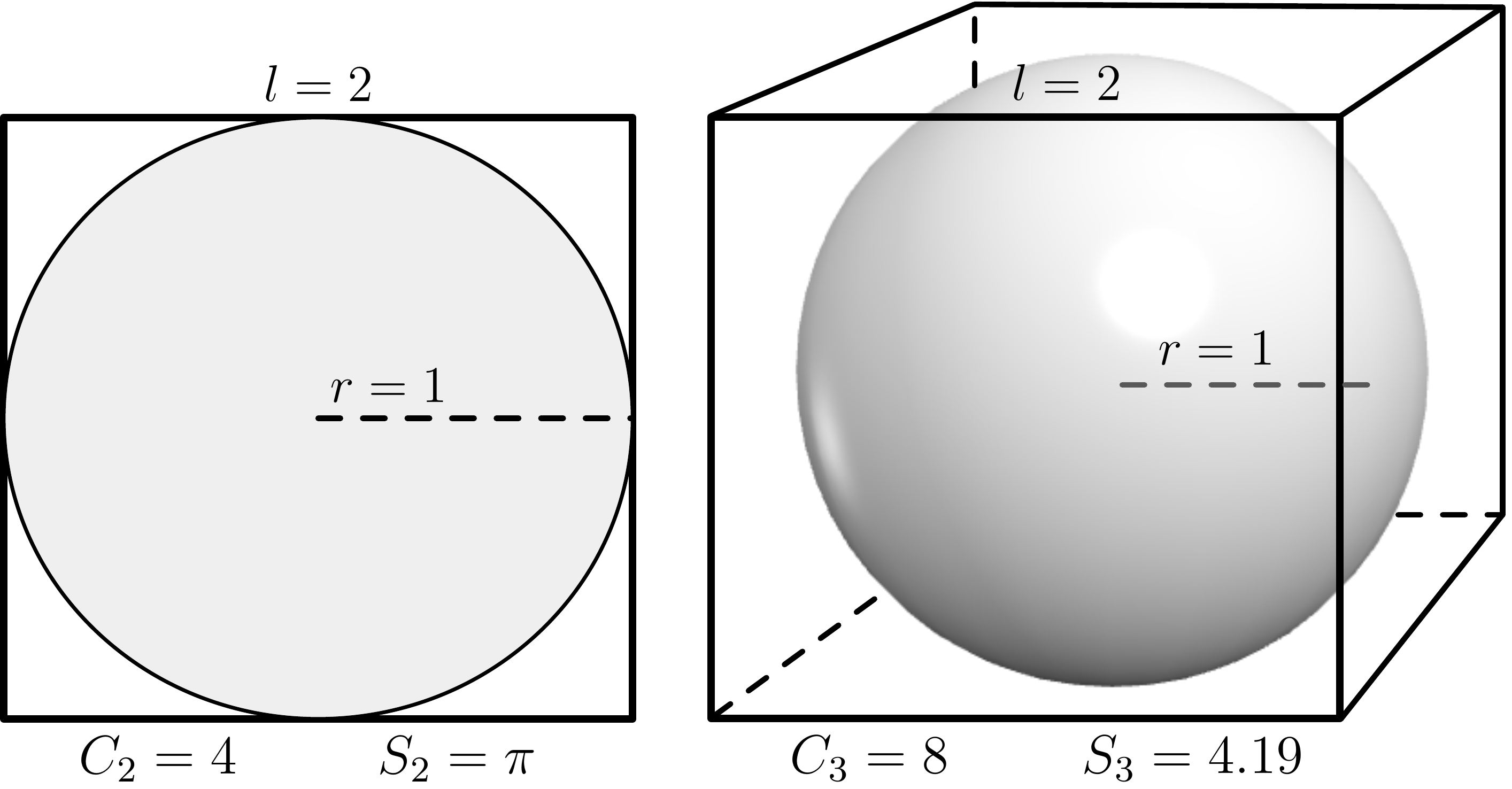}
\newline
\caption{Depicts a circle inscribed within a square (left) and a sphere inscribed within a cube (right). Even though the diameters of the circle and the sphere are the same as the lengths of the sides of the square and the cube, the ratio between the volume of the circle and the volume of the square  greatly decreases when moving from two to three dimensions.}
\label{fig:circs}
\end{figure}

Finally, the peculiarities of high-dimensional space are well visualized geometrically. Figure \ref{fig:circs} shows the generalization of a circle in 2-$D$ inscribed within a square, to a sphere in 3-D inscribed within a cube. Taking this further still, the cube and the sphere can be generalized to hyper-cubes and hyper-spheres, the volumes for which can be calculated as $C_D = l^D $ and $S_D= \frac{\pi^{D/2}}{\Gamma\left( \frac{D}{2}+1\right)}$, respectively. The latter is a generalization of the well-known expression for the volume of a sphere. The link with the previous discussion lies in the fact that Gaussian data are spherically (or at least elliptically) distributed. As such, an exploration of the characteristics of spheres and ellipses in high-dimensions tells us something about about high-dimensional data.

\begin{figure}[!t]
\centering
\includegraphics[scale=0.3]{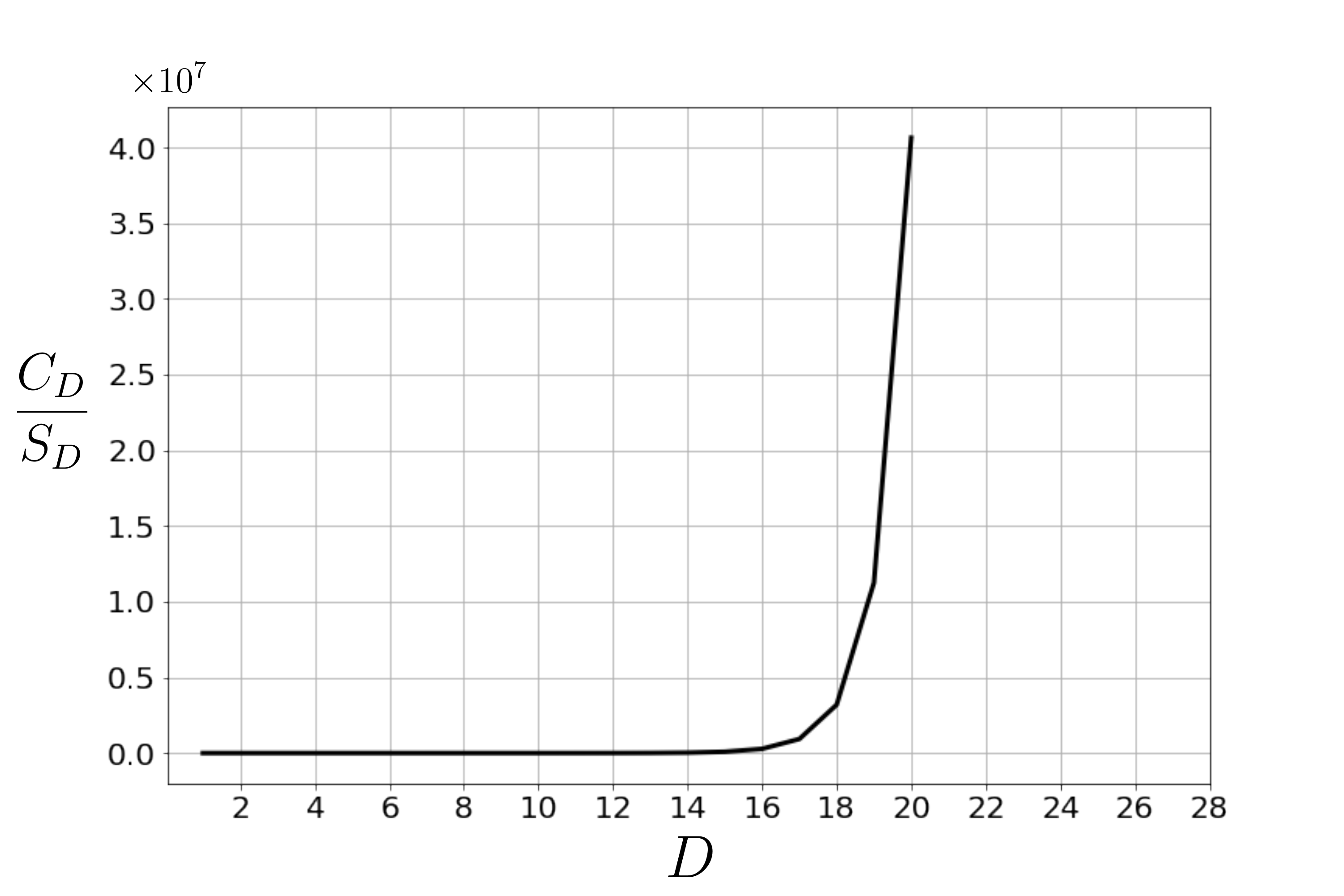}
\newline
\caption{Depicts the ratio between the volume of a hupercube and the volume of a hypersphere as the dimensionality $D$ increases. Note the scale of the $y$-axis ($\times 10^7$). }
\label{fig:ratios}
\end{figure}

\begin{figure}[!t]
\centering
\includegraphics[scale=0.3]{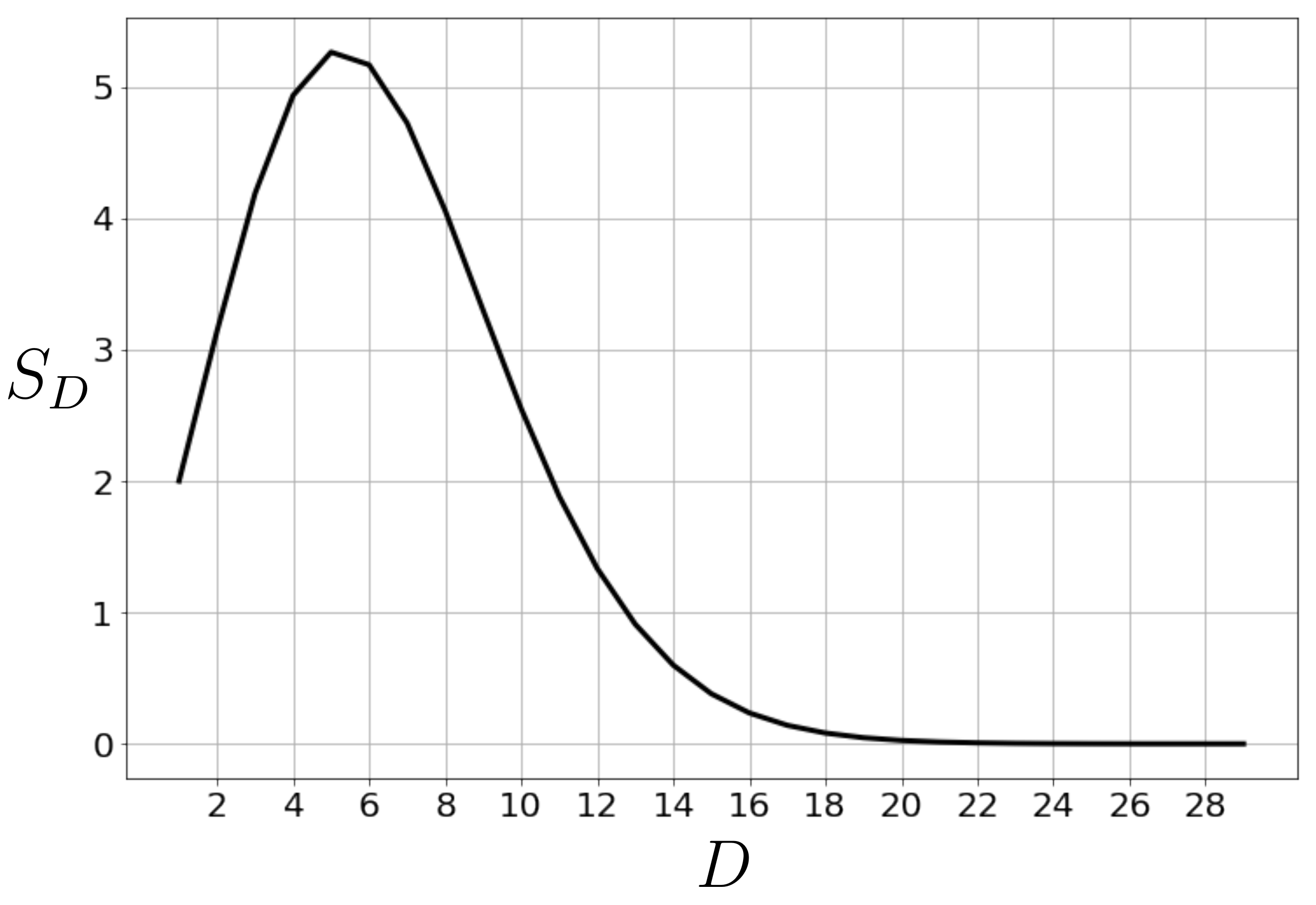}
\newline
\caption{Shows how the volume $S_D$ of a hypersphere changes with dimensionality $D$. }
\label{fig:spherevol}
\end{figure}

Figure \ref{fig:circs} illustrates how the ratio between the volume of a sphere and the volume of a cube changes dramatically even though the number of dimensions has only increased by 1. In the first case, considering the square and the circle, the ratio between the volumes is $4/\pi \approx 1.27$. In the second case, considering now the cube and the sphere, the ratio between the volumes is $8/4.19 = 1.90$. In other words, the volume of the sphere represents a significantly smaller fraction of the total volume of the cube even though only one extra dimension has been added. Figure \ref{fig:ratios} illustrates how this pattern continues exponentially, such that the ratio between a cube and sphere for $D=20$ is over 40 Million. In other words, a cube with sides of length two has a volume which is 40 million times greater than the sphere inscribed within it. This effect is at least partly explained by the extraordinary way the volume of a sphere changes as dimensionality increases. Figure \ref{fig:spherevol} shows how the volume of a sphere with a radius of one in $D$-dimensions quickly tends to zero after passing a maximum at around $D=5$. In other words, a high-dimensional sphere \textit{has negligible volume}.

Whilst the implications of this are the same as those in the previous section, the demonstration hopefully gives some further intuition about just how quickly the strange effects start to occur as $D$ is increased, at least in the case where our dimensions are independent. In order to gain an intuition for whether these effects translate to more realistic data (including correlated dimensions/variables), see the analysis below. Many problems in social science involve more than just a few dimensions, and problems which utilise `big data' are even more susceptible to issues relating to what is known as the curse of dimensionality.

\section{Typicality: An Information Theoretic Way to Characterize `Normality'}
In the previous section, we described how randomly sampled vectors in high-dimensional space tend to be located at a radius of length $\sqrt{D}$ away from the mean, and tend to be uncorrelated. This makes points close to the mean across multiple dimensions poor examples of `normality'. In this section we introduce the concept of \textit{typicality} from information theory, as a means to categorize whether a particular sample or a particular set of samples is/are `normal' or `abnormal' (and therefore also whether the points should be considered to be outliers). 

\subsection{Entropy}

Entropy describes the degree of surprise, uncertainty, or information associated with a distribution and is computed as $-\sum_{i=1}^N p(x_i) \log p(x_i)$ where $N$ is the number of datapoints in the sample distribution, $x_i$ is a single datapoint in this distribution, and $p(x_i)$ is that datapoint's corresponding probability.\footnote{We temporarily consider the discrete random variable case for this example, but note that the intuition holds for continuous distributions as well.} If the entropy is low, it means the distribution is more certain and therefore also easier to predict.

Taking a fair coin as an example, $p(x=\mbox{heads}) = p(x=\mbox{tails}) = 0.5$. The entropy of this distribution is $H = -(0.5\log_2 0.5 + 0.5 \log_2 0.5) = 1$. Recall from above that entropy describes the amount of information content - the units of entropy here are in bits. The fact that our fair coin has 1 bit of information should therefore seem quite reasonable - there are two equally possible outcomes and therefore one bit's worth of information. Furthermore, because the coin is unbiased, we are unable to predict the outcome any better than by randomly guessing. On the other hand, let's say we have a highly biased coin whereby $p(x=\mbox{heads}) = 0.99$ and $p(x=\mbox{tails}) = 0.01$. In this case $H = -(0.99\log_2 0.99 + 0.01 \log_2 0.01) = 0.08$. The second example had a much lower entropy because we are likely to observe heads, and this makes samples from the distribution more predictable. As such, there is less new or surprising information associated with samples from this distribution, than there was for the case where there was an equal chance of a head or a tail.

According to the Asymptotic Equipartition Property, entropy can be approximated as a sum of log probabilities of a sequence of random samples, in a manner equivalent to how an expected value can be estimated as a sum of random samples according to the Law of Large Numbers \cite{cover2006}:

\begin{equation}
\begin{split}
    \frac{-1}{N} \log p(\mathbf{x}_{i=1}, \mathbf{x}_2, ..., \mathbf{x}_N) \approx  \\ \frac{-1}{N} \sum_{i=1}^N \log p(\mathbf{x}_i) \approx \\ H(\mathbf{x})  \mbox{ for sufficiently large }N
    \end{split}
    \label{eq:AEP}
\end{equation}

In words, the negative log of the joint probability tends towards the entropy of the distribution. Entropy therefore gives us an alternative way to characterize normality; but now instead of doing so using the arithmetic mean, we do so in terms of entropy. Rather than comparing the value of a new sample against the mean or expected value of a distribution, we can now consider the probability of observing that sample and its relation to the entropy of the distribution.

\subsection{Defining the Typical Set}
We are now ready to define the typical set. Rather than comparing datapoints against the mean, we can compare them against the entropy of the distribution $H$. For a chosen threshold $\epsilon$, datapoints may be considered typical according to \cite{Dieleman2020, cover2006, MacKay2018}:

\begin{equation}
    \mathcal{T} = \{\mathbf{x}: 2^{-(H+\epsilon)} \leq p(\mathbf{x}) \leq 2^{-(H-\epsilon)} \}
    \label{eq:typicalset}
\end{equation}

In words, the typical set $\mathcal{T}$ comprises datapoints $\mathbf{x}$ which fall within the bounds defined on either side of the entropy of the distribution. Datapoints which have a probability close (where close is defined according to the magnitude of $\epsilon$) to the entropy of the distribution are thereby defined as typical. Recall the thin annulus containing most of the probability mass, illustrated in Figure \ref{fig:Mackay}; this annulus comprises the typical set. Note that, because this annulus contains most of our probability mass, the set quickly incorporates all datapoints as $\epsilon$ is increased \cite{cover2006}. Note that this typical set (at least for `modest' values of $\epsilon$) does not contain the mean because, as an annulus, it cannot contain it by design (the mean falls at the centre of a circle who's radius defines the radius of the annulus). The quantity given in Eq.~\ref{eq:typicalset} can be computed for continuous Gaussian (rather than discrete) data using the analytical forms for entropy $H$ for the univariate and multivariate Gaussian provided as Supplementary, and the probability density function for a univariate or multivariate Gaussian for $p(\mathbf{x})$. This is undertaken for the outlier detection simulation below.

\subsection{Establishing Typicality in Practice}

Even though it is arguable as to whether the Gaussian should be used less ubiquitously for modeling data distributions than it currently is \cite{Micceri1989}, one of the strong advantages of the Gaussian is its mathematical tractability. This tractability enables us to calculate (as opposed to estimate) quantities exactly, simply by substituting parameter values into the equations (assuming these parameters have themselves not been estimated). Thus, moving from a comparison of dataset values against the average or expected value to a consideration for typicality does not necessitate the abandonment of convenient analytic solutions. A derivation of the (differential) entropy for a Gaussian distribution has been provided in supplementary material, and is given in Eq. \ref{eq:entgauss}.

\begin{equation}
H(f) = \frac{1}{2}\log_2(2\pi e\sigma^2)
\label{eq:entgauss}
\end{equation}

Note that the mean does not feature in Eq. \ref{eq:entgauss} - this make it clear that the uncertainty or information content of a distribution is independent of its location (\textit{i.e.}, the mean) in vector space.\footnote{Note that entropy is closely related to the score function (the derivative of the log likelihood) as well as Fisher information, which is the variance of the score.} As well as being useful in categorising datapoints as typical or atypical (or, alternatively, inliers and outliers) in practice, Eq. \ref{eq:entgauss} can also be used to understand the relationship between $\epsilon$ and the fraction of the total probability mass that falls inside the typical set. Returning to Figure \ref{fig:typicals} which shows the lengths of 2,000 vectors sampled from a 40-dimensional Gaussian, we can see that as $\epsilon$ increases, we gradually expand the interval to cover a greater and greater proportion of the empirical distribution. Note also that the mean, which in this plot has a location (0,0), is a long way from any of the points and is not part of (and, by definition, cannot be part of) the typical set.


\section{An Example with Real-World Data}
To demonstrate that these effects do not only apply to idealistic simulations, we use the LISS longitudinal panel data, which is open access \cite{Scherpenzeel2010}. Specifically, we use Likert-style response data from wave 1 of the Politics and Values survey, collected between 2007 and 2008, which includes questions relating to levels of satisfaction and confidence in science, healthcare, the economy, democracy, etc. Given that no inference was required for these data, a simple approach was taken to clean it: all non-Likert style data were removed, leaving 58 variables, and text based responses which represented the extremes of the scale we replaced with integers (\textit{e.g.}, `no confidence at all' is replaced with a 0). For the sake of demonstration, all missing values were mean-imputed\footnote{Across all included variables the amount of mean-imputation, on average, was 7.9\%. Note that such imputation makes the demonstration more conservative, because it forces values to be equal to the mean for the respective dimension.} (this may not be a wise choice in practice), and the data were standardized so that all variables were mean zero with a standard deviation of one. In total there were 6,811 respondents.

\begin{figure}[!t]
\centering
\includegraphics[scale=0.4]{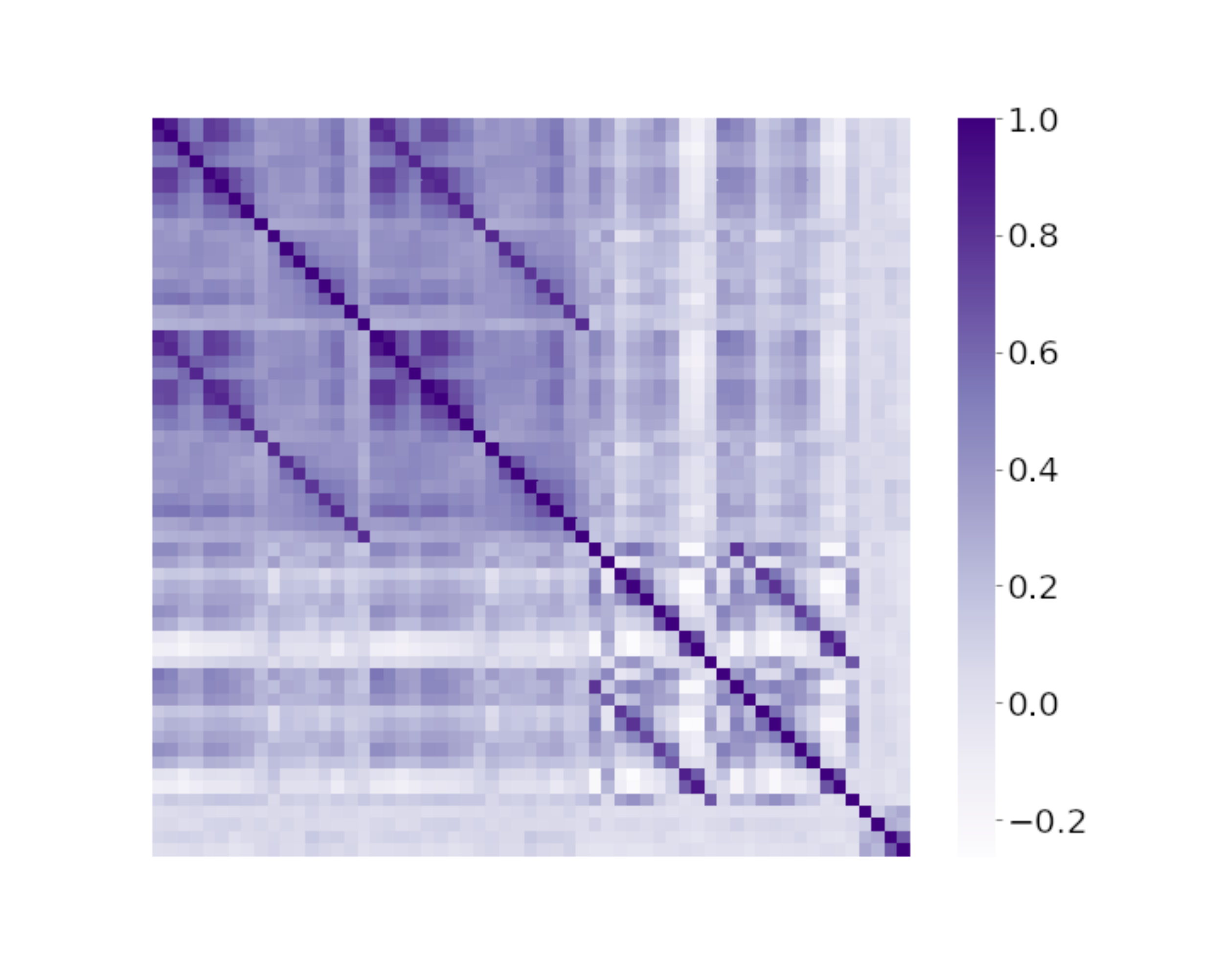}
\newline
\caption{Depicts the bivariate correlations for the LISS panel data \cite{Scherpenzeel2010}.}
\label{fig:correlations}
\end{figure}

\begin{figure}[!t]
\centering
\includegraphics[scale=0.3]{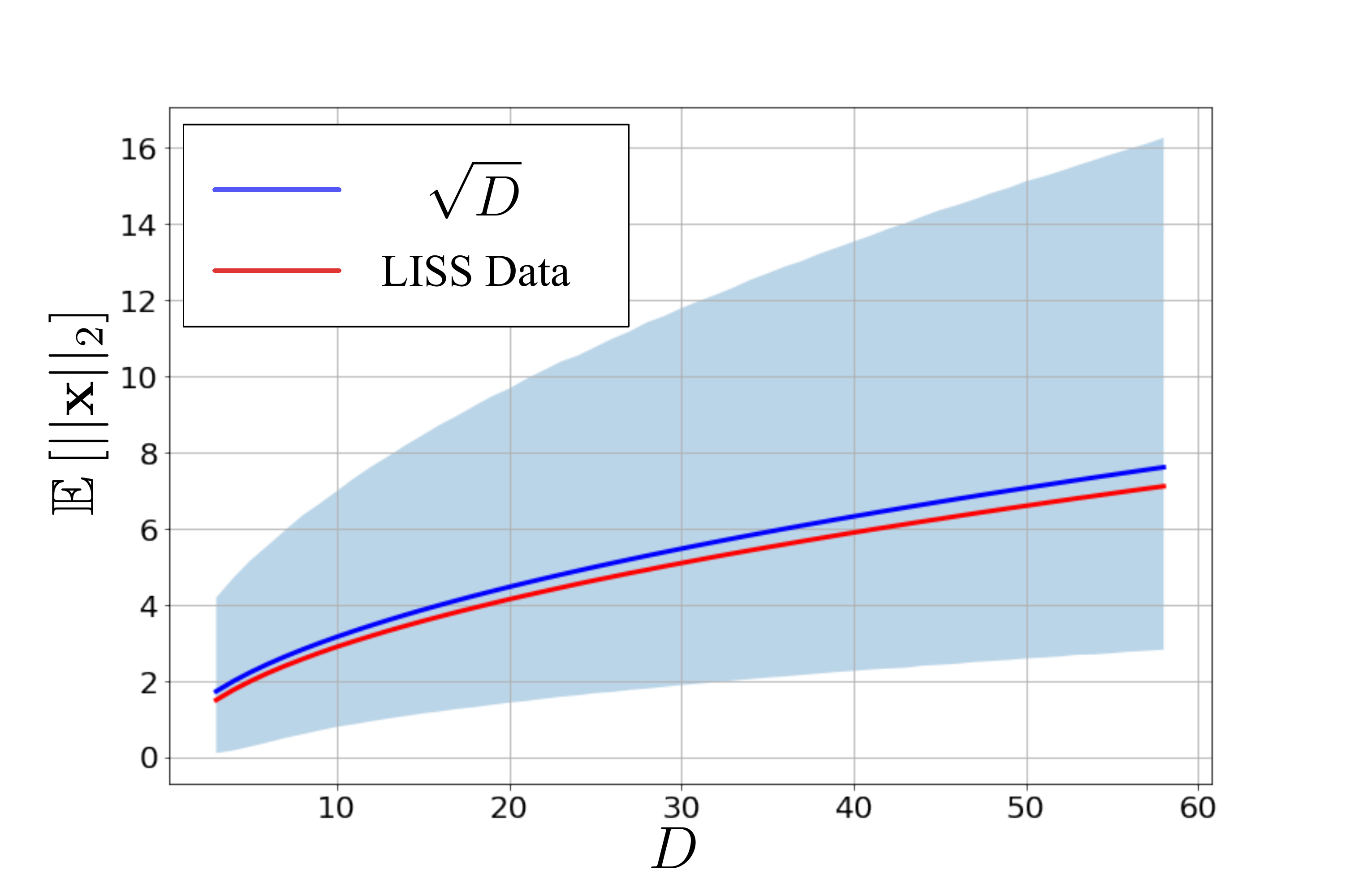}
\newline
\caption{The lengths for vectors from the LISS panel data (red), for increasing $D$, as well as the expected lengths for a multi-variate Gaussian (blue). The LISS panel data curve includes 1-99\% percentile intervals  \cite{Scherpenzeel2010}.}
\label{fig:lisslengths}
\end{figure}

Figure \ref{fig:correlations} depicts the bivariate correlations for each pair of variables in the data. It can be seen that there exist many non-zero correlations, which makes these data useful in understanding the generality of our expositions above (which were undertaken with independent and therefore uncorrelated variables). Qualitatively, some variables were also highly non-Gaussian, which again helps us understand the generality of the effects in multi-dimensional data. Figure \ref{fig:lisslengths} shows how the expected lengths of the vectors in the LISS panel data change as an increasing number of dimensions are used. To generate this plot, we randomly selected $D$ variables 1000 times, where $D$ range from three up to the total number of variables (58). For each of the 1,000 repetitions, we computed the Euclidean distances of each vector in the dataset across these $D$ variables, and then computed their average. Once the 1,000 repetitions were complete, we compute the average across these repetitions to obtain an approximation to the expectation of vector lengths in $D$ dimensions. Finally, we overlaid a plot of $\sqrt{D}$ to ascertain how close the empirically estimated vector lengths are, compared with the expected lengths for a multivariate Gaussian. We also plot the 1-99\% intervals, which are found to be quite wide, owing to the mix of lowly and highly correlated variables in conjunction with possibly non-Gaussianity.

These results demonstrate that even for correlated, potentially non-Gaussian, real-world data, the peculiar behavior of multi-dimensional data discussed in this paper nonetheless occur. For the LISS data, the expected lengths were slightly lower than for samples from a `clean' multivariate Gaussian, and this is likely to be due to the correlations present in the data.\footnote{For further discussion relating to this point, see Kroc \& Astivia \cite{Kroc2021}.} Indeed, when estimating the entropy of these data (and as can be seen in the supplementary code), a robust covariance matrix estimation approach was used to account for the non-isotropic nature of the joint distribution. More generally, non-parametric methods can be used for the estimation of typicality, but such methods are likely to have lower sample efficiency (i.e., more data are required for accurate estimation of entropy and typicality).

\section{Moving Forward with Multivariate Outlier Detection}

Grubbs defined outliers as samples which ``deviate markedly from other members of the sample in which it occurs'' \cite{Grubbs1969}. This definition is useful to us here, because it is not expressed in terms of distance from the mean, but in broad/general terms. Indeed, as we have already discussed, in as few as four dimensions, points near the mean become increasingly unlikely. This suggests that outlier methods should not only identify points which are too far from the mean, but also those which are too close. 

Two related definitions of outliers which were noted by Leys et al. \cite{LeysOutliers} are: ``Data values that are unusually large or small compared to the other values of the same construct'', and ``Data points with large residual values.'' The first is quite similar to Grubbs' definition, identifying values as unusually large or small (\textit{i.e.}, deviating markedly) with respect to other values of the same construct (\textit{i.e.}, with respect to the other members of the sample in which they occur). The second defines them with respect to the residuals of a statistical model. In other words, they are values which lead to large discrepancies between true and predicted values. Note that both of these definitions bear the consequences for our work - whether we are comparing datapoints against the rest of the sample, or comparing them against the predictions from a statistical model designed to estimate an expected value (which is by far the most common case in psychology and social science), the relevance of these definitions to our discussion remains the same.

It is also, perhaps, of interest to note that our definition of outliers makes no value judgement about whether outliers are good or bad. Indeed, depending on the application and our research questions, outliers may represent `golden' samples. Consider a manufacturer interested in fabricating the perfect mechanical prototype. Each sample may have its own unique blemishes, and our target may represent the perfect average across all (high-dimensional) opportunities for such blemishes. In such a case, the average represents the golden target for our manufacturer, and identifying it necessitates outlier detection methods which understand that values across high-dimensions close to the mean should be considered to be (in this case, desirable) outliers, in much the same way as samples which deviate because they are too far from the mean may also be outliers for opposite reasons.

Leys et al. \cite{LeysOutliers} provide a useful summary of options for both univariate and multivariate outlier detection, as well as a discussion about the consequences of outlier management decisions.  Whilst their work provide an excellent introduction to multivariate outlier detection and good practice, they do not discuss the strange behavior of the mean in multiple dimensions, nor the impact of this behavior on multivariate outlier detection methods which are unable to detect outliers which lie close to the mean

We note, as other researchers have \cite{LeysOutliers}, that the most common method used for multidimensional/multivariate outlier detection in the domain of social science is the Mahalanobis distance \cite{Mahalanobis1930}. For a description of the Mahalanobis distance and its application, readers are directed to work by Li et al. \cite{LiOutlier} and Leys et al. \cite{LeysOutlier2018}. Briefly, the method assesses the distance of a point from the centroid (\textit{i.e.}, the mean) of a cloud of points in (possibly correlated) multidimensional space. The researchers note that in order to compute the distance from putative outliers to the mean, it is first necessary to estimate the mean and covariance whilst including those points in the estimation \cite{LeysOutliers2013, LeysOutlier2018}. This process is somewhat problematic because if outliers are included in the calculation being used to compute the mean and covariance, the estimation of these quantities will themselves be biased towards these outliers, thereby reducing the chances of correctly identifying the outliers. A solution is proposed which is called the `robust' Mahalanobis distance \cite{LiOutlier, LeysOutlier2018}, which leverages what is known as the Minimum Covariance Determinant (MCD), and estimates the centroid / mean by selecting an estimate of the mean from a set of estimates derived from different subsets of the dataset. 

Unfortunately, despite the Mahalanobis distance and its robust variant being the most commonly used multidimensional/multivariate outlier detection techniques in social science, it suffers from the same problems as any multidimensional method based on distances from the centroid/mean. By consequence it would certainly not flag someone average in all dimensions as an outlier, even though statistically they would represent an extremely unusual individual (it would not help Daniels with his project, for example). It is therefore important that researchers qualify their definition of the outlying set to explicitly admit points which may fall too close to the mean.

When using the Mahalanobis distance, one can make decisions about the set of outliers $\mathcal{O}$ using the following expression:

\begin{equation}
    \mathcal{O} = \{\mathbf{x} :  M(\mathbf{x}) > c \},
    \label{eq:mahalanobis}
\end{equation}

where $ M(\mathbf{x}) = \sqrt{(\mathbf{x} - \hat{\boldsymbol{\mu}})^T \mathbf{S}^{-1}(\mathbf{x}  -\hat{\boldsymbol{\mu}})}$ and is the estimated Mahalanobis distance (in units of standard deviation) for the multivariate datapoint under consideration $\mathbf{x}$, and $c$ is the threshold for classifying a point as an outlier. In the expression for $M$, $\hat{\boldsymbol{\mu}}$ is the estimate of the mean of the distribution, $\mathbf{S}$ is the estimated covariance matrix. One of the benefits of the Mahalanobis based methods is that one can use them to threshold the data based on units of standard deviations. Thinking in terms of standard deviations is not unusual and therefore the process of selecting outliers in these terms thus leads to intuitive selection thresholds. In contrast, we see in Eq.~\ref{eq:typicalset} that the threshold for determining whether a datapoint falls within the typical set $\mathcal{T}$ depends on $\epsilon$, which is not related to the standard deviation, but rather to a distance away from the entropy.

\begin{figure}[!t]
\centering
\includegraphics[scale=0.35]{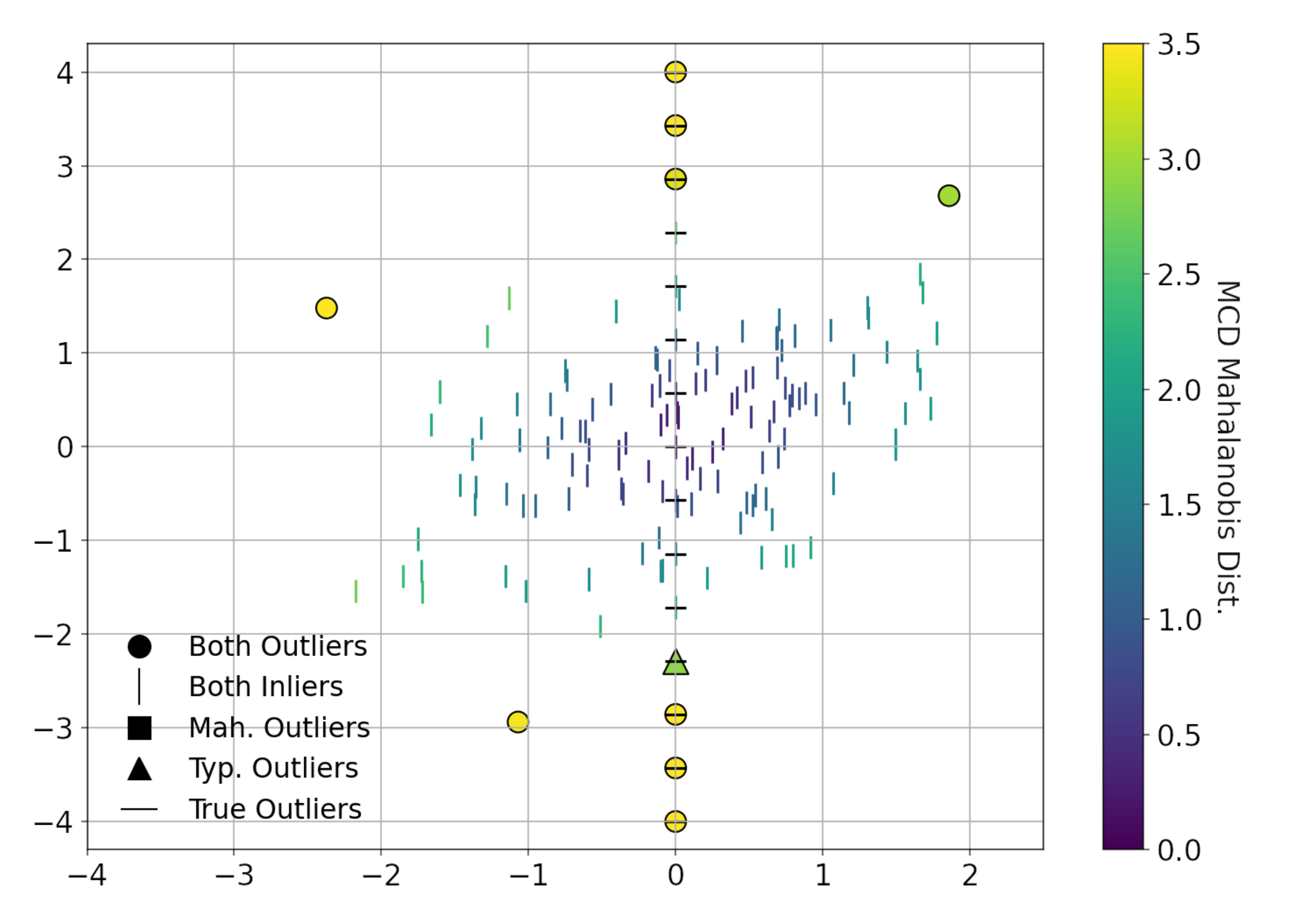}
\newline
\caption{Comparison of Mahalanobis distance and typicality for outlier detection. The outliers are generated as a vertical set of equally spaced points (indicated with horizontal dashes) ranging from negative four to positive four on the y-axis, superimposed on a set of 125 points (indicated with vertical dashes) drawn from a bivariate Gaussian with a covariance of 0.5. The points identified to be outliers by both methods are indicated in circles, whilst those indicated to be outliers by the the Mahalanobis or typicality methods separately are indicated by squares or triangles, respectively. The color of the points represents the Mahalanobis distance in units of standard deviation. The estimation of the covariance matrices for both methods used the robust Minimum Covariance Determinant (MCD) method. Note that there are no squares because no points were uniquely detected as outliers according to Mahalanobis distance.}
\label{fig:outliercomparison}
\end{figure}

We have already seen how typicality has the added advantage of classifying datapoints which lie too \textit{close} to the mean. In Figure~\ref{fig:outliercomparison} we show that, in low-dimensional settings, typicality can be used to make approximately the same classification of outliers as the Mahalanobis distance to the extent that some datapoints which lie \textit{far} from the mean should still be classified as outliers. Of course, in practice a balance must be struck between the value of $\epsilon$ in Eq.~\ref{eq:typicalset}, in the same way that $c$ in Eq.~\ref{eq:mahalanobis} must be decided. 

Specifically, for Figure~\ref{fig:outliercomparison}, we generated 125 points from a bivariate Gaussian with a covariance of 0.5, and then added a set of equally spaced outlier points ranging from negative four to positive four on the y-axis (indicated with horizontal dashes). As such, not all these points are expected to be identified as outliers, because some of their values lie well within the tails of the distribution. They do, however, enable us to compare at which point they are identified as outliers by the two detection methods under comparison. Note that the subsequent estimation is done after the creation of the complete dataset (including the outliers) using all the empirical values. Using the robust MCD estimator mentioned above, we computed both the Mahalanobis distance (in units of standard deviation), and colored each point according to this distance. For typicality, we followed the estimation of entropy for the multivariate Gaussian which also takes in an estimate for the covariance (see the Supplementary Material for the relationship between the covariance matrix and the entropy of a Gaussian), for which we again used the MCD method. The use of MCD for typicality arguably arguably makes our typicality estimator `robust' for the same reason that it is considered to make the Mahalanobis distance estimation robust. The threshold for the Mahalanobis distance was set to three standard deviations, whilst the value for the typicality threshold was set to five. In practice, researchers may, of course, need to suitably select and justify these values. 

The scatter-plot marker shapes are set according to whether the outliers were classified as such by both methods (circles), just the Mahalanobis method (squares), or just the typicality method (triangles). If neither method classifies a point as an outlier, the points are set to vertical dashes (\textit{i.e.}, `inliers'). Note that there are no points which are classified as outliers by the Mahalanobis method which are not also classified as outliers by the typicality method. The inverse is not quite true, with one additional point (indicated with the triangle marker) being classified as an outlier by the typicality method.\footnote{Although this classification is correct, this point lies on the limit of the cloud of true inliers, and so in practice it would not be clear whether this would represent a useful outlier classification or not.} Figure~\ref{fig:outliercomparison} therefore indicates that, in low-dimensions, Mahalanobis distance performs similarly to typicality as an outlier detection method.

\begin{figure}[!t]
\centering
\includegraphics[scale=0.35]{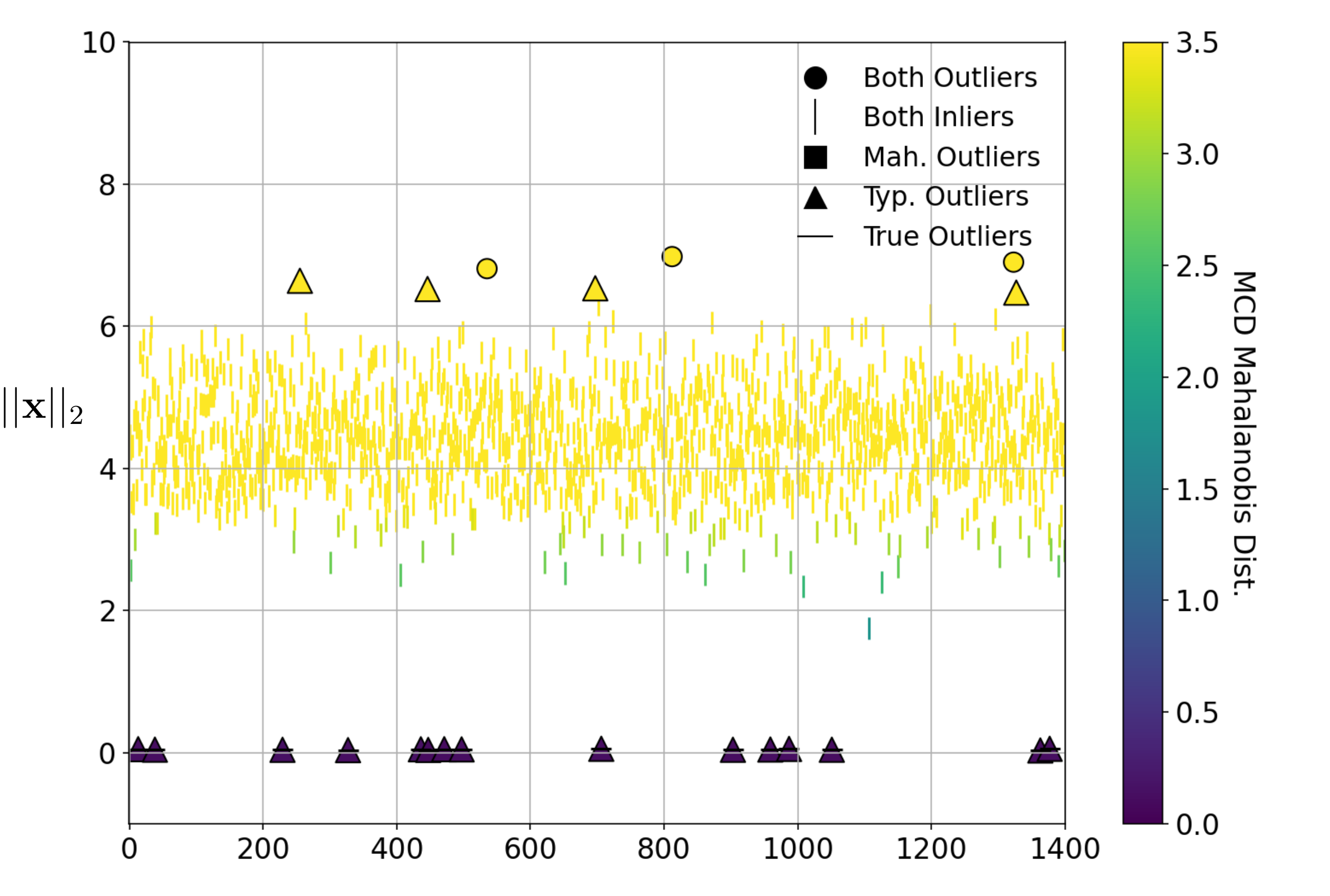}
\newline
\caption{Comparison of Mahalanobis distance and typicality for outlier detection in 20 dimensional space. The outliers are generated as a set of 15 points close to the expected value of 0, superimposed on a set of 1400 points drawn from a 20 dimensional isotropic Gaussian. For visualisation purposes, this plot shows the lengths of each point (the x-axis is simply the index of the point in the dataset). The points identified to be outliers by both methods are indicated in circles, whilst those indicated to be outliers by the the Mahalanobis or typicality methods separately are indicated by squares or triangles, respectively. The color of the points represents the Mahalanobis distance in units of standard deviation. The estimation of the covariance matrices for both methods used the robust Minimum Covariance Determinant (MCD) method. Note that there are no squares because no points were uniquely detected as outliers according to Mahalanobis distance}
\label{fig:outliercomparison2}
\end{figure}

In Figure~\ref{fig:outliercomparison2}, we undertake the same task, but this time in a 20-dimensional space. The figure shows the lengths of each of 1400 points (the lengths are used for visualisation purposes) drawn from a 20-dimensional, isotropic Gaussian. Fifteen of these points are manually set to fall very close to the mean / expected value of zero, and these are the simulated outliers we wish to identify which fall towards the bottom of the plot. Now, in contrast to the example above, we see a large difference between the outliers identified using the two methods. Typicality successfully identifies all 15 true outliers as outliers, whereas MCD fails to identify \textit{any} of them. Conversely, some points which lie far from the mean (but which have a low probability of occurrence relative to the entropy of the distribution) are identified by both MCD and typicality, although it is possible that by tweaking the thresholds one could achieve greater overlap between the classification of these points by the two methods.

In summary, typicality does not only have a role in detecting outliers in high-dimensional scenarios (where the outliers may include values close to the expected value), but can perform similarly to how current approaches (such as MCD) do in low-dimensional scenarios, which otherwise fail in high-dimensions. We thus recommend practitioners consider typicality as a viable outlier detection approach under both low- and high-dimensional conditions, and especially in high-dimensions. To this extent, researchers are encouraged to consult various commentaries on the usage of outlier detection methods, such as the one by Leys et al. \cite{LeysOutliers} which provides general recommendations for practice (including pre-registration). It is notable that prior commentary does not include a discussion about the limitations of Mahalanobis based methods for outlier detection once the number of dimensions increases, which serves as a reminder of how important it is that researchers explore typicality. We recommend updating the working conceptualisation of outliers to include those points which, in high-dimensions (but as few as 4-10 dimensions) fall too close to the mean. 

Finally, we note a quite different approach to identifying outliers known as cellwise-outlier detection \cite{Raymaekers2021} which can be used to identify outliers at a more granular level (identifying not only which cases are outliers, but which variables are responsible for this classification). Note, however, that this approach does not include a discussion about the additional complications that arise as dimensionality increases (specifically, a discussion about how the relevance of identifying unusually high or low values for certain variables might change if one considers that values close to the average are atypical). Further work is required to more broadly evaluate the implications of the atypicality of the mean across other statistical approaches. 

\section{Conclusion}
The principal goal of this work was to provide researchers with an intuition about the behavior of data as the number of dimensions increases. Through our exploration of multi-dimensional space, we have shown that the mean, far from representing normality, actually represents abnormality, in so far as encountering a datapoint close to the mean in datasets comprising more than a handful of dimensions becomes incredibly unlikely, even with a large number of datapoints. In contrast with the arithmetic average, the information theoretic quantity known as  `typicality' provides a way to establish normality (or rather, whether a datapoint is typical or atypical), which is particularly useful in high-dimensional regimes. Given that researchers in psychology and social science frequently deal with multivariate datasets, and that the peculiarities associated with multi-dimensional spaces start occurring in relatively low dimensions (as few as four), it is important that researchers have some awareness of the concepts presented in this paper. Indeed, the implications of this work are particularly important to consider for researchers concerned with policy. Policies, especially in areas like social welfare, health, and education, are often built around statistics related to `average' individuals. If the representational relevance of the mean is limited, particularly in multi-dimensional contexts, then policy design, implementation, and evaluation decisions based on this perspective may be misdirected. 

Clearly, the motivations behind the characterizations of points as either normal or abnormal overlap strongly with those behind outlier detection. The discussion also provides us with a good justification for updating our working definition of `outlier' to include points which lie unusually close to the mean. Unlike popular multivariate outlier detection techniques such as the Mahalanobis distance, which characterize outliers as points which lie \textit{far} from the expected value of the distribution, typicality additionally offers a means to detect those which are \textit{close}. Whilst such additional benefits of typicality based methods become more evident as the dimensionality of the dataset increases (where traditional methods like Mahalanobis distance fail) we showed that typicality also performs as one would hope/expect in low dimensions. To show this, we finished with an evaluation of typicality for bivariate outlier detection using a `robust' version of entropy using the Minimum Covariance Determinant estimation technique, and verified via simulation that in low-dimensions it works well as an alternative to the popular Mahalanobis distance.

In addition, it is worth noting that we used a closed-form, parametric expression for entropy and for the probability of individual datapoints $p(\mathbf{x})$ (parametric in the sense that we assumed an underlying multivariate Gaussian with a covariance and mean). Such a parametric approach carries the advantage of high sample efficiency, that is, relatively few datapoints are required to adequately estimate the relevant quantities. However, in practice one may (a) have more dimensions than datapoints, (b) have non-Gaussian data, or (c) both. When the number of datapoints is low compared with the number of dimensions, parametric estimators may exhibit substantial instability (especially the estimator for the covariance). In the case where the data cannot be justifiably parameterised by a Gaussian (or, indeed, any other parametric distribution), semi- or non-parametric approaches can be used to estimate the entropy and $p(\mathbf{x})$. Unfortunately, there are concomitant disadvantages regarding sample efficiency. In other words, the number of samples required for reliable estimation goes up considerably, and this requirement is disproportionately problematic for data with high-dimensionality. The extent to which this is a problem depends on the combination of the specific choice of estimators, the sample size, and the dimensionality of the data. A detailed exploration of the interplay between these factors is beyond the scope of this work, and it should be mentioned that the point is important regardless of whether one is intending to detect outliers, or undertaking statistical modeling of high-dimensional data in general.

\section{Acknowledgements}
I'd like to direct interested readers to the excellent blog posts by Sander Dieleman \cite{Dieleman2020} and Stefan Stein \cite{Stein2020} , which both partly inspired this work, and provided much food for thought. Thanks also to CentERdata for access to their LISS panel data (\url{https://www.lissdata.nl/}).

{\small
\bibliographystyle{ieee_fullname}
\bibliography{NN}
}

\begin{appendices}
\section{Differential Entropy of a Gaussian}

Following Cover \& Thomas \cite{cover2006}, the \textit{differential entropy} (in bits) is defined as:

\begin{equation}
    H(f) = -\mathbb{E}[\log(f(x))] = - \int_{-\infty}^{+\infty}f(x) \log_e f(x) dx 
\end{equation}

The probability density function of the normal distribution is:
\begin{equation}
    f(x) = \frac{1}{\sqrt{2 \pi \sigma^2}}e^{-\frac{(x-\mu)^2}{2\sigma^2}}
\end{equation}

Substituting the expression for $f(x)$ into $h(f)$:

\begin{equation}
H(f) = -\int_{-\infty}^{+\infty}f(x) \log_e \left[\frac{1}{\sqrt{2 \pi \sigma^2}}e^{-\frac{(x-\mu)^2}{2\sigma^2}} \right]
\end{equation}

\begin{equation}
\begin{split}
H(f) = \\ -\int_{-\infty}^{+\infty}f(x) \log_2e\left(\log_e \left(\frac{1}{\sqrt{2\pi \sigma^2}} \right) + \log_e e^{-\frac{(x-\mu)^2}{2\sigma^2}} \right)   
\end{split}
\end{equation}

\begin{equation}
\begin{split}
H(f) = \\ -\int_{-\infty}^{+\infty}f(x) \log_2e\left(-\log_e (\sqrt{2\pi\sigma^2)} - \frac{(x-\mu)^2}{2\sigma^2}  \right)
\end{split}
\end{equation}

\begin{equation}
\begin{split}
H(f) = \\ \log_2 e \log_e \sqrt{2\pi \sigma^2}\int_{-\infty}^{+\infty}f(x)dx + \\ \log_2e\int_{-\infty}^{+\infty} \frac{(x-\mu)^2}{2\sigma^2}f(x) dx
\end{split}
\end{equation}

Note that:
\begin{equation}
\int_{-\infty}^{+\infty}f(x)dx =1
\end{equation}

and recall that:

\begin{equation}
  \int_{-\infty}^{+\infty} (x-\mu)^2f(x) dx = \mathbb{E}[(x-\mu)^2] = \mbox{Var}(x) = \sigma^2
\end{equation}

Therefore: 
\begin{equation}
H(f) = \log_2 \sqrt{2\pi\sigma^2} + \frac{\log_2e}{2\sigma^2} \sigma^2
\end{equation}

And finally:

\begin{equation}
H(f) = \frac{1}{2}\log_2(2\pi e\sigma^2)
\end{equation}

For a $D$-dimensional Gaussian, the derivation for the entropy is as follows:

\begin{equation}
H(f) = -\mathbb{E}[\log(f(\mathbf{x}))] = - \int_{-\infty}^{+\infty}f(\mathbf{x}) \log_e f(\mathbf{x}) d\mathbf{x} , 
\end{equation}

where the bold font indicates multidimensionality.

\begin{equation}
\begin{split}
H(f) = \\ -\mathbb{E}[\log[(2\pi)^{-D/2} |\mathbf{S}|^{-0.5} \mbox{exp}(-0.5(\mathbf{x} - \boldsymbol{\mu})^T \mathbf{S}^{-1}(\mathbf{x} - \boldsymbol{\mu}))]],
\end{split}
\end{equation}

where $\mathbf{S}$ is the covariance matrix, $T$ indicates the transpose, and $|.|$ indicates the determinant. 

\begin{equation}
\begin{split}
H(f) =\\ 0.5D \log(2\pi) + 0.5 \log|\mathbf{S}| + 0.5 \mathbb{E}[(\mathbf{x} - \boldsymbol{\mu})^T \mathbf{S}^{-1}(\mathbf{x} - \boldsymbol{\mu})]
\end{split}
\end{equation}

\begin{equation}
 = 0.5D (1 + \log(2\pi)) + 0.5\log(\mathbf{S})
\end{equation}

This last expression can then be expressed in bits by multiplying by $\log_2 e$. Note that this derivation follows the approach provided by \cite{Gundersen2020} and uses a number of `tricks' relating to the trace operator.

\end{appendices}
\end{document}